\documentclass[doublecol]{epl2}

\usepackage{amsmath}
\usepackage{color}

\title{Thermodynamic competition between membrane protein\\ oligomeric states}
\shorttitle{Thermodynamic competition between membrane protein oligomeric states} 

\author{Osman Kahraman \and Christoph A. Haselwandter}
\shortauthor{Osman Kahraman and Christoph A. Haselwandter}

\institute{Department of Physics \& Astronomy and Molecular and Computational Biology Program, \\Department of Biological Sciences, University of Southern California, Los Angeles, CA 90089, USA}


\abstract{Self-assembly of protein monomers into distinct membrane protein oligomers provides a general mechanism for diversity in the molecular architectures, and resulting biological functions, of membrane proteins. We develop a general physical framework describing the thermodynamic competition between different oligomeric states of membrane proteins. Using the mechanosensitive channel of large conductance as a model system, we show how the dominant oligomeric states of membrane proteins emerge from the interplay of protein concentration in the cell membrane, protein-induced lipid bilayer deformations, and direct monomer-monomer interactions. Our results suggest general physical mechanisms and principles underlying regulation of protein function via control of membrane protein oligomeric state.}

\begin{document}

\maketitle

\section{Introduction}
\begin{figure}
\onefigure{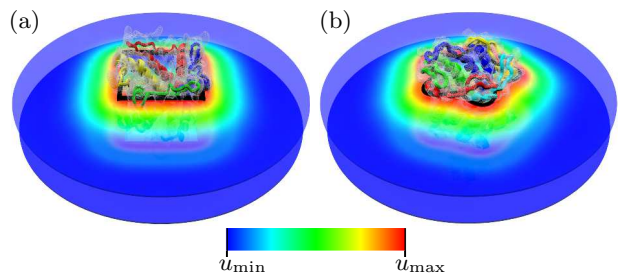}
\caption{The observed (a) tetrameric \cite{liu09} and (b) pentameric \cite{chang98} MscL structures result \cite{plos13,epl14,scirep16} in distinct lipid bilayer deformation footprints. The protein-induced lipid bilayer thickness deformations in (a) and (b) were calculated by minimizing eq.~(\ref{eq:elastic_energy}) using finite elements \cite{epl14,scirep16,pre16}. The color scale ranges from $u_\text{min} = 0$ to $u_\text{max} = 0.3$~nm. The black curves in (a) and (b) illustrate the polygonal and clover-leaf models of the shape of multimeric membrane proteins considered here \cite{plos13,epl14}, which are motivated by the observed tetrameric \cite{liu09} and pentameric \cite{chang98} MscL structures, with $\alpha=0.22$ for the clover-leaf model. (Structural data shown as ribbon diagrams; Protein Data Bank accession numbers 3HZQ and 2OAR for tetrameric and pentameric MscL, respectively.)
}
\label{fig1}
\end{figure}

Membrane proteins perform a wide variety of biological functions, which requires
\cite{vinothkumar2010,forrest15} a diverse array of molecular architectures of membrane proteins. Structural biology has shown \cite{goodsell2000,forrest15,takamori06,yun11,linden12} that cells often achieve diversity in membrane protein architecture through self-assembly of small protein subunits into membrane protein oligomers. Intriguingly, a range of experiments suggest \cite{gandhi11,reading15,walton15} that a given membrane protein may exist in more than one oligomeric state (quaternary structure), which is expected \cite{vinothkumar2010,forrest15,goodsell2000} to affect its biological function. In particular, structural studies indicate \cite{walton15} that ion channels which show large conformational changes during gating tend to be composed of monomer subunits that allow multiple oligomeric states. These general observations are exemplified \cite{gandhi11,reading15,walton15} by the mechanosensitive channel of large conductance (MscL) \cite{perozo03,booth07,kung10,haswell11}, which is gated by membrane tension and provides a model system for mechanosensation. Protein crystallography has yielded tetrameric \cite{liu09} (see fig.~\ref{fig1}(a)) and pentameric \cite{chang98,sukharev01a,sukharev01b,walton13} (see fig.~\ref{fig1}(b)) MscL structures, while electron microscopy experiments have suggested \cite{saint98} that MscL is a hexamer. The oligomeric state of MscL has also been studied using a variety of specialized biophysical and biochemical techniques \cite{sukharev99,gandhi11,reading15,walton15,dorwart10,iscla11}. The physiologically relevant oligomeric states of MscL remain a matter of debate \cite{sukharev99,gandhi11,reading15,walton15,dorwart10,iscla11,haswell11}, but available data suggests \cite{walton15} that MscL may occur as a mixture of different oligomeric states \textit{in vivo}, with pentameric MscL being predominant.

While the oligomeric state of a membrane protein can be critical for its
biological function \cite{vinothkumar2010,goodsell2000,forrest15,takamori06,yun11,linden12},
the physical mechanisms controlling membrane protein oligomeric state remain largely unknown. In this letter we develop, using MscL in the closed state as a model system, a general physical framework describing the thermodynamic competition between different oligomeric states of membrane proteins. Based on \textit{in vitro} experiments on MscL \cite{reading15} we thereby assume that membrane proteins can interconvert, and attain equilibrium, between specific oligomeric states. Following previous work on the statistical thermodynamics of amphiphile \cite{benshaul1994,safran2003} and viral capsid \cite{bruinsma03} self-assembly, we construct a simple free energy describing self-assembly of membrane protein oligomers. This free energy involves two key contributions stemming from the internal energy of membrane protein oligomers. On the one hand, membrane proteins deform the surrounding lipid bilayer (fig.~\ref{fig1}), yielding a lipid bilayer deformation energy characteristic of the overall shape of membrane proteins. The resulting energy per protein is, in general, not proportional to the number of monomers forming a given membrane protein
oligomer \cite{mondal13,plos13,epl14,scirep16}, and hence affects the competition between different oligomeric states. On the other hand, the energy cost of direct protein-protein interactions between membrane protein subunits \cite{vinothkumar2010,forrest15,goodsell2000,gandhi11,walton15} may change with protein oligomeric state, which can also bias membrane protein self-assembly towards specific oligomeric states. Finally, we consider the effect of protein clustering via lipid-bilayer-mediated interactions \cite{harroun99,goforth03,botelho06,phillips09,grage11,nomura12} on the thermodynamic competition between membrane protein oligomeric states. Our model shows how the dominant oligomeric states of membrane proteins emerge from the interplay of protein concentration in the cell membrane, protein-induced lipid bilayer deformations, and direct monomer-monomer interactions. Our results suggest general physical mechanisms and principles underlying regulation of protein function via \cite{vinothkumar2010,goodsell2000,forrest15,takamori06,yun11,linden12,gandhi11,reading15,walton15}
control of membrane protein oligomeric state.

\section{Thermodynamic model}
We formulate a thermodynamic model of the competition between different oligomeric
states of membrane proteins following a formalism developed previously in
the context of amphiphile \cite{benshaul1994,safran2003} and viral capsid \cite{bruinsma03} self-assembly in dilute aqueous solutions. Our starting point is to assume \cite{benshaul1994,safran2003,bruinsma03} that the entropy of the system can be approximated by the mixing entropy of proteins and solvent molecules in the membrane,
\begin{equation} \label{eq:entropy}
S=-k_B\sum_s \frac{N_s}{s} \left(\text{ln} \frac{N_s}{s N_l}-1\right) ,
\end{equation}
where $k_B$ is Boltzmann's constant, $N_s$ denotes the total number of monomers bound in oligomers with $s$ monomers each, and $N_l$ denotes the total number of solvent molecules, which we take to be dominated by contributions due to lipids.

In eq.~(\ref{eq:entropy}) it is assumed that membrane protein oligomers do not interact with each other, a point we return to below. Furthermore, eq.~(\ref{eq:entropy}) assumes that the total monomer number fraction $M=\sum_s N_s/N_l \ll 1$. We confirm the validity of this assumption for MscL in \textit{E. coli} using a sphero-cylindrical model of the shape of \textit{E. coli} \cite{phillips12} with an overall cell length $\approx 2$~$\mu$m and radius of curvature $\approx0.5$~$\mu$m at the cell poles. Assuming that approximately half of the membrane area is covered by lipids \cite{linden12}, this yields $N_l\approx4.9\times10^6$ for a lipid radius $\approx0.45$~nm \cite{damodaran93}. The protein copy number in the cell membrane $N_P \approx200$--$1100$ for MscL in \textit{E. coli} under physiological conditions \cite{bialecka12} which, under the assumption \cite{walton15,reading15} that MscL mostly occur as tetramers, pentamers, or hexamers in \textit{E. coli}, yields $M\approx1.6\times10^{-4}$--$1.3\times10^{-3}\ll 1$. While in these simple estimates we have focused on MscL in \textit{E. coli}, similar considerations are expected to apply to other membrane proteins and organisms.

Equation~(\ref{eq:entropy}) yields the Helmholtz free energy \cite{benshaul1994,safran2003,bruinsma03}
\begin{align}
F=\sum_s\left[N_s \epsilon_s +k_B T \frac{N_s}{s} \left(\ln \frac{N_s}{s N_l}-1 \right) \right]
 \label{eq:free}
\end{align}
for a mixture of membrane protein oligomeric states in the dilute regime, where $T$ is the temperature and $\epsilon_s$ is the energy per monomer in oligomers with $s$ subunits each. Minimization of the free energy in eq.~(\ref{eq:free}) with respect to $N_s$ subject to the constraint $\sum_s N_s/N_l=M$,
which imposes a fixed total monomer number fraction in the membrane,~yields
\begin{align} \label{eq:distribut}
\epsilon_s+ \frac{k_B T}{s}\ln \frac{N_s}{s N_l} =\mu \,,
\end{align}
where $\mu$ is the Lagrange multiplier associated with the constraint $\sum_s N_s/N_l=M$ and can be interpreted as a chemical potential \cite{benshaul1994,safran2003,bruinsma03}.
For fixed $\epsilon_s$ and $M$, eq.~(\ref{eq:distribut}) provides a set of equations that we solve numerically to determine the fraction of monomers in a given oligomeric state, $\phi_s=N_s/\sum_s N_s$, in thermal equilibrium.

Equation~(\ref{eq:distribut}) shows that any contribution to $\epsilon_s$
that does not change with $s$ only corresponds to a shift in $\mu$, and hence
does not affect the thermodynamic competition between different oligomeric
states. We focus here on two distinct sets of contributions to $\epsilon_s$. On the one hand, the energy of direct protein-protein interactions between the monomers forming an oligomer may depend on $s$, yielding contributions to $\epsilon_s$ that vary with $s$. We return to such contributions to $\epsilon_s$ below. On the other hand, $\epsilon_s$ is also expected to involve contributions due to interactions between the membrane protein and the surrounding lipid bilayer. Indeed, bilayer-protein interactions have been found to regulate membrane protein function \cite{jensen04,lundbaek06,andersen07,phillips09,mcintosh06,brown12,mouritsen93,engelman05},
and to play a fundamental role in orchestrating membrane protein oligomerization and sculpting the shape of membrane proteins \cite{white01,booth05,anbazhagan10,bogdanov14}. In the standard elasticity theory of bilayer-protein interactions \cite{jensen04,lundbaek06,andersen07,phillips09,pre16}, integral membrane proteins are modeled as rigid membrane inclusions that deform the surrounding lipid bilayer \cite{boal02,safran2003,seifert97}, yielding protein-induced lipid bilayer curvature \cite{canham70,helfrich73,evans74} and thickness \cite{huang86} deformations. For closed MscL, structural data \cite{chang98} suggests \cite{wiggins04,wiggins05} that contributions to the energetic cost of bilayer-protein interactions due to protein-induced bilayer curvature deformations are at least one order of magnitude smaller than contributions due to protein-induced
bilayer thickness deformations. We therefore focus here on the lipid bilayer thickness deformation energy \cite{huang86,andersen07,wiggins04,wiggins05,ursell08}
\begin{align} 
G=\frac{1}{2}\int \text{d}x \text{d}y \left[K_b \left(\nabla^2 u\right)^2+K_t \left(\frac{u}{a}\right)^2 \right] ,
\label{eq:elastic_energy}
\end{align}
where the thickness deformation field $u(x,y)$ is one-half the protein-induced perturbation in bilayer hydrophobic thickness, $K_b$ is the bending rigidity, $K_t$ is the thickness deformation modulus, $a$ is one-half the hydrophobic thickness of the unperturbed lipid bilayer, and, for simplicity, we have set the membrane tension equal to zero. The values of $K_b$, $K_t$, and $a$ generally depend on the lipid composition \cite{rawicz00}. Following previous work on bilayer-protein
interactions for MscL and other membrane proteins \cite{phillips09, andersen07}
we use here $K_b=20$~$k_B T$ and $K_t=60$~$k_B T/\text{nm}^2$, and a value $a=1.6$~nm suggested by x-ray scattering experiments on the \textit{E. coli} cytoplasmic membrane \cite{mitra04}. Models based on eq.~(\ref{eq:elastic_energy}) have been found to capture the basic phenomenology of bilayer-protein interactions in a variety of different experimental model systems \cite{andersen07,phillips09,jensen04,lundbaek06,mcintosh06,brown12,
huang86,helfrich90,nielsen98,nielsen00,harroun99b,partenskii02,partenskii03,partenskii04,kim12,lundbaek10,greisen11,wiggins04,wiggins05,ursell07,ursell08,grage11,plos13,epl13,epl14,scirep16,mondal11,mondal12,mondal13,mondal14,plos14}.

We determine contributions to $\epsilon_s$ due to bilayer-protein interactions by minimizing eq.~(\ref{eq:elastic_energy}) with respect to $u$ subject to a given protein symmetry (oligomeric state) and shape, and corresponding boundary conditions on $u$. Inspired by the observed structures of tetrameric \cite{liu09} and pentameric \cite{chang98} MscL, we allow here
for two families of protein cross section \cite{plos13,epl14}: polygonal and clover-leaf shapes. In particular, the structure of tetrameric MscL \cite{liu09} suggests a tetragonal shape (fig.~\ref{fig1}(a)) while the structure of pentameric MscL \cite{chang98} suggests a five-fold-symmetric clover-leaf shape (fig.~\ref{fig1}(b)).
We implement polygonal shapes as described previously \cite{plos13}
using a cross-sectional area $A_s=s A_m$ of membrane proteins, where the monomer area $A_m\approx3.32$~nm$^2$ is estimated \cite{plos13} from the structure of closed pentameric MscL \cite{chang98}. Clover-leaf shapes are defined by the boundary curve~\cite{plos13,epl14}
\begin{align} \label{eq:clover}
C_s(\theta) = R\left(1+\alpha \cos s \theta \right)
\end{align}
in polar coordinates with the protein center at the origin of the coordinate
system, in which $R$ sets the protein size and $\alpha$ captures the amplitude of angular undulations. Similarly as for polygonal protein shapes, we fix $R$ by demanding that $A_s=s A_m$. Unless indicated
otherwise, we use the values $\alpha=0.22$ and $7.1\times10^{-2}$ for MscL pentamers and hexamers estimated \cite{plos13} from the structure of pentameric MscL \cite{chang98} and electron micrographs of hexameric MscL \cite{saint98}, respectively. We use a hydrophobic thickness $=3.8$~nm of MscL as suggested \cite{ursell08,plos13} by structural studies \cite{chang98,elmore03}. Following previous work \cite{ursell07,wiggins04,ursell08,wiggins05,nielsen98,huang86}
we employ zero-slope boundary conditions at the bilayer-protein interface and assume that thickness deformations decay away from the membrane proteins.
Equation~(\ref{eq:elastic_energy}) can be minimized for arbitrary protein symmetries using \cite{pre16} analytic \cite{plos13,plos14,epl13} or numerical
\cite{epl14,scirep16} approaches. We use the finite element approach described
in refs.~\cite{pre16,epl14,scirep16}. While we focus here on MscL as a model system, available data on membrane protein structure \cite{vinothkumar2010,forrest15} suggests that the polygonal and clover-leaf models considered here may also
provide coarse-grained representations of the shapes of other multimeric membrane proteins.

\begin{figure}
\includegraphics[width=0.98\columnwidth]{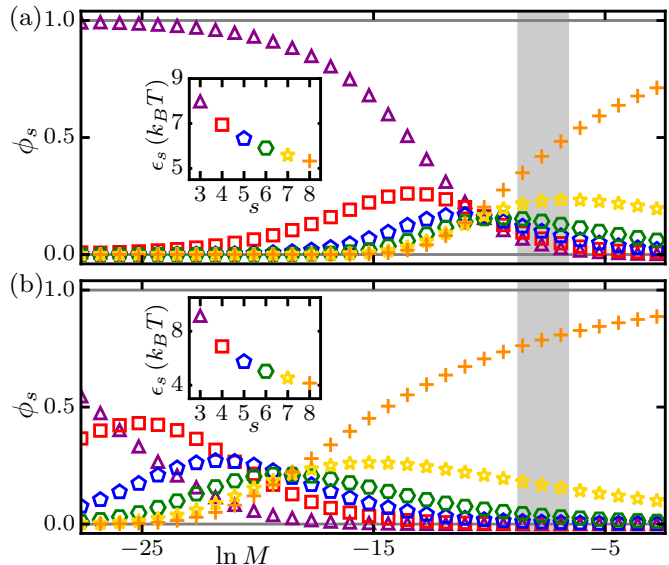}
\caption{Monomer fractions $\phi_s$ in oligomeric states $s=3,\dots,8$ \textit{vs.} total monomer number fraction in the membrane, $M$, obtained from eq.~(\ref{eq:distribut}) for (a) clover-leaf shapes with $\alpha=0.22$ and (b) polygonal shapes, with the corresponding $\epsilon_s$, calculated from the bilayer thickness
deformation energy in eq.~(\ref{eq:elastic_energy}), shown in the insets. For reference, the range $M=1.6\times10^{-4}$--$1.3\times10^{-3}$ associated with MscL in \textit{E. coli} under physiological conditions \cite{bialecka12}
is indicated by grey areas.
}
\label{fig2}
\end{figure}


\section{Bilayer-protein interactions}
Allowing only for contributions to $\epsilon_s$ due to bilayer-protein interactions
we first consider the thermodynamic competition between different (non-interacting) oligomeric
states of clover-leaf (see fig.~\ref{fig2}(a)) and polygonal (see fig.~\ref{fig2}(b)) membrane protein shapes as a function of total monomer number fraction, $M$. In
particular, we set $\epsilon_s=G_s/s$, where $G_s$ is computed for each $s$ and membrane protein shape from eq.~(\ref{eq:elastic_energy}) using finite elements. We focus on $s$ ranging from $s=3$ (trimers) to $s=8$ (octamers). Independent of the particular shape considered, we find that small $M$ favor low-symmetry oligomers while large $M$ favor high-symmetry oligomers.  This can be understood \cite{safran2003}
from eq.~(\ref{eq:distribut}) by noting that $\mu$ increases with $M$, with  $\mu<\text{min}(\epsilon_s)$ since we assume here $M \ll 1$. From eq.~(\ref{eq:distribut}) we also have 
\begin{equation}
N_s=s N_l e^{-s\left(\epsilon_s-\mu\right)/k_B T} \,,
\end{equation}
implying for $\mu \ll \text{min}(\epsilon_s)$ (small $M$) that the monomer
fraction $\phi_s$ is small for large $s$. However, the lipid bilayer contributions to $\epsilon_s$ decrease with $s$ for clover-leaf as well as polygonal shapes (see fig.~\ref{fig2}(a), inset, and fig.~\ref{fig2}(b), inset), making higher-order symmetries favorable as far as bilayer-protein interactions are concerned. For sufficiently large $M$ and $s$, the magnitude of $s(\epsilon_s-\mu)$ can therefore become small enough to permit the highly symmetric oligomeric states dominating the large-$M$ regimes in fig.~\ref{fig2}, with entropic effects being suppressed \cite{safran2003}. For intermediate $M$, we find coexistence of all allowed oligomer symmetries at approximately equal concentrations
(fig.~\ref{fig2}(a)) or a hierarchy of dominating oligomer symmetries of increasing $s$ with increasing $M$ (fig.~\ref{fig2}(b)) depending on the particular model of protein shape considered.

\begin{figure}
\includegraphics[width=0.98\columnwidth]{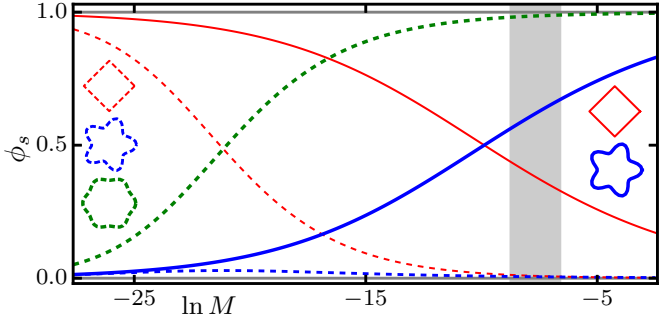}
\caption{Monomer fractions $\phi_s$ obtained from eq.~(\ref{eq:distribut}) for tetrameric \cite{liu09} and pentameric \cite{chang98} MscL (solid curves), and tetrameric \cite{liu09}, pentameric \cite{chang98}, and hexameric \cite{saint98} MscL (dashed curves), \textit{vs.} total monomer number fraction in the membrane, $M$. The range $M=1.6\times10^{-4}$--$1.3\times10^{-3}$ corresponding to MscL in \textit{E. coli} under physiological conditions \cite{bialecka12} is indicated by a grey area. All monomer energies were calculated from the bilayer thickness deformation energy in eq.~(\ref{eq:elastic_energy}) using MscL shapes suggested by protein crystallography \cite{chang98,liu09} and electron microscopy \cite{saint98} as discussed in ref.~\cite{plos13} (see insets), and are given by $\epsilon_{4,5,6}\approx 6.88$, $6.32$,~$5.01$~$k_B T$. }
\label{fig3}
\end{figure}

Protein crystallography has yielded a tetrameric structure of truncated MscL
in \textit{S. aureus} \cite{liu09}, which resembles a tetragonal shape (fig.~\ref{fig1}(a)), and a pentameric structure of MscL in \textit{M. tuberculosis} \cite{chang98}, which resembles a five-fold clover-leaf shape (fig.~\ref{fig1}(b)). Assuming
\cite{reading15} that MscL subunits may form either this tetrameric or pentameric structure in the cell membrane we find, consistent with the results in fig.~\ref{fig2}, that tetrameric MscL dominates at small $M$ but pentameric MscL dominates at large $M$ (see fig.~\ref{fig3}). Furthermore, we find that, for the range of values of $M$ thought to be relevant for \textit{E. coli} under physiological conditions \cite{bialecka12}, pentameric MscL dominates over tetrameric MscL, consistent with experiments \cite{sukharev99,gandhi11,reading15,walton15,dorwart10,iscla11,haswell11}
on the oligomeric state of MscL.

Electron microscopy has suggested a hexameric clover-leaf shape of MscL in \textit{E. coli} \cite{saint98}. It is unclear \cite{sukharev99,gandhi11,reading15,walton15,dorwart10,iscla11,haswell11},
however, whether the resolution in these electron microscopy experiments was sufficient to allow unambiguous assignment of the oligomeric state of MscL. Indeed, protein crystallography has yielded \cite{walton13} a pentameric structure of the cytoplasmic domain of MscL in \textit{E. coli}. Allowing for thermodynamic competition between tetrameric, pentameric, and hexameric MscL, we find that tetrameric MscL dominates at small $M$ and hexameric MscL dominates at large $M$, with pentameric MscL being effectively suppressed for the entire $M$-range considered here (fig.~\ref{fig3}). This can be understood intuitively by noting that the aforementioned electron microscopy experiments on MscL in \textit{E. coli} \cite{saint98} suggest a hexameric clover-leaf
shape of MscL with, compared to the observed pentameric structure of MscL in \textit{M. tuberculosis} \cite{chang98}, small $\alpha$ \cite{plos13}, yielding only small deviations from a circular protein cross section (see eq.~(\ref{eq:clover})). Thus, bilayer-protein interactions strongly penalize \cite{plos13} pentameric MscL compared to hexameric MscL for the pentameric and hexameric MscL shapes considered here \cite{saint98,chang98}, effectively suppressing pentameric MscL.


\section{Monomer-monomer interactions}
One expects that, for $s>2$, contributions to the protein energy due to direct protein-protein interactions between the monomers forming an oligomer are, to leading order, proportional to $s$. As discussed above, such constant contributions to $\epsilon_s$ do not affect the thermodynamic competition between different oligomeric states of membrane proteins. In general, however, protein-protein interactions can also yield contributions to $\epsilon_s$ that vary with $s$. In the case of MscL, for instance, different oligomeric states imply \cite{gandhi11,walton15} different crossing angles between the $\alpha$-helices in neighboring MscL subunits, which may modify the direct protein-protein interaction energy between MscL monomers. To account for such contributions to $\epsilon_s$ we write
\begin{align}
 \epsilon_s = \frac{1}{s} \left(G_s + E_P\right) ,
\label{eq:nonbilayer}
\end{align}
where, as above, $G_s$ is computed for each $s$ and membrane protein shape from eq.~(\ref{eq:elastic_energy}) and $E_P$ models contributions to $\epsilon_s$ due to direct protein-protein interactions between the monomers forming an oligomer.

\begin{figure}
\includegraphics[width=0.98\columnwidth]{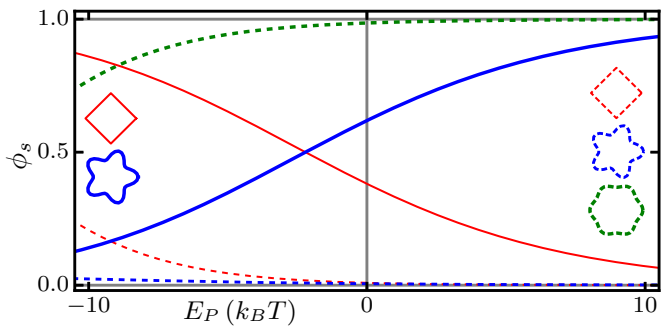}
\caption{Monomer fractions $\phi_s$ obtained from eq.~(\ref{eq:distribut}) with eq.~(\ref{eq:nonbilayer}) for tetrameric \cite{liu09} and pentameric \cite{chang98} MscL (solid curves), and tetrameric \cite{liu09}, pentameric \cite{chang98}, and hexameric \cite{saint98} MscL (dashed curves), \textit{vs.} $E_P$ using a total monomer number fraction $M=4.7\times10^{-4}$ corresponding to MscL in \textit{E. coli} under physiological conditions \cite{bialecka12}. The contributions to $\epsilon_s$ due to bilayer-protein interactions, $G_s/s$,
are the same as in fig.~\ref{fig3}.}
\label{fig4}
\end{figure}

In general, $E_P$ in eq.~(\ref{eq:nonbilayer}) may vary with $s$ but, for simplicity, we take here $E_P$ to be a constant. This can be visualized as $s \to \infty$ serving as a reference state, with decreasing $s$ yielding increasingly large corrections, parameterized by $E_P$, to the monomer energy. For generality we allow for $E_P>0$ as well as $E_P<0$, with an approximate energy scale $|E_P|\leq10$~$k_B T$ \cite{white01} for interactions between protein helices in a bilayer environment. Specific estimates of $E_P$ could potentially be obtained via molecular dynamics simulations.
Note that $E_P<0$ favors oligomeric states with fewer monomers, while $E_P>0$ favors oligomeric states with more monomers. Using a total monomer number fraction $M=4.7\times10^{-4}$ corresponding to MscL in \textit{E. coli} under physiological conditions \cite{bialecka12}, we find that the main results in fig.~\ref{fig3} are robust with respect to $s$-dependent variations in the protein-protein interaction energy between monomers (see fig.~\ref{fig4}). In particular, allowing for thermodynamic competition between tetrameric, pentameric, and hexameric MscL, with the MscL shapes suggested \cite{plos13} by structural studies \cite{chang98,liu09} and electron microscopy \cite{saint98}, we find that hexameric MscL dominates for the entire $E_P$-range considered here. Furthermore, we find that, if monomers can only form tetrameric \cite{liu09} or pentameric \cite{chang98} MscL, pentameric MscL dominates for $E_P>0$ as well as $E_P<0$ provided that the magnitude of $E_P$ is not too large, while tetrameric MscL dominates for $E_P<0$ and large magnitudes of $E_P$. The latter result follows because, as already mentioned above, $E_P<0$ in eq.~(\ref{eq:nonbilayer}) penalizes oligomeric states with large $s$.

\section{Bilayer-mediated protein clustering}
\begin{figure}
\center
\includegraphics[width=0.98\columnwidth]{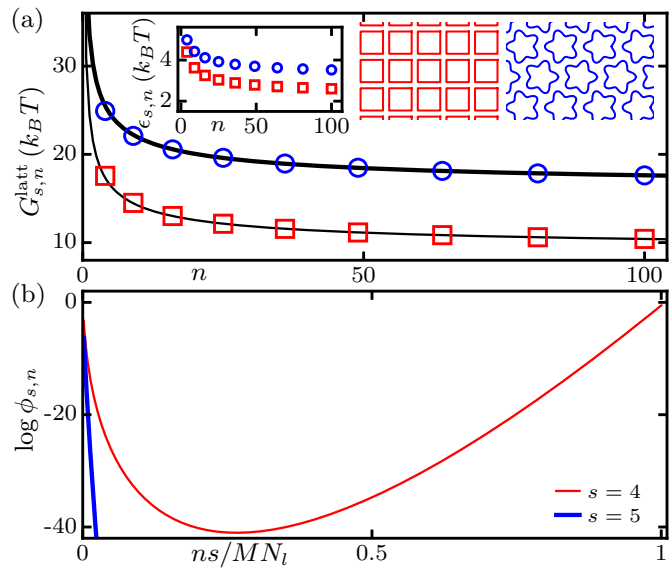}
\caption{Bilayer-mediated protein clustering. (a) Lipid bilayer thickness deformation energy in eq.~(\ref{eq:elastic_energy}) per MscL, $G_{s,n}^{\text{latt}}$,
and per MscL monomer, $\epsilon_{s,n}$, (left inset) \textit{vs.} number of MscL in the lattice, $n$, computed using finite elements \cite{epl14,scirep16,pre16} for face-on square (squares) and distorted hexagonal \cite{henley86} (circles) lattice architectures \cite{scirep16} of tetrameric \cite{liu09} and pentameric \cite{chang98} MscL clusters (central and right insets) with $\sqrt{n}$ MscL along the horizontal and vertical directions, and corresponding fits to eq.~(\ref{eq:lattice}) (black curves) with $G^{\infty}_4=8.68$~$k_B T$ and $G^{\prime}_4=17.31$~$k_B T$, and $G^{\infty}_5=15.66$~$k_B T$ and $G^{\prime}_5=19.64$~$k_B T$. (b) Monomer fractions $\phi_{s,n}$ for tetrameric \cite{liu09} and pentameric \cite{chang98} MscL \textit{vs.} scaled lattice size $ns/M N_l$, obtained from eq.~(\ref{eq:EqCluster}) with eq.~(\ref{eq:lattice}) for the face-on square (thin red curve) and distorted hexagonal \cite{henley86} (thick blue curve) lattice architectures of tetrameric and pentameric MscL clusters in (a) \cite{scirep16} using a total monomer number fraction $M=4.7\times10^{-4}$ corresponding to MscL in \textit{E. coli} under physiological conditions \cite{bialecka12}.
}
\label{fig5}
\end{figure}

So far in this letter we have focused on individual, non-interacting membrane proteins. For membrane proteins in close enough proximity, protein-induced
lipid bilayer deformations overlap. The resulting bilayer-mediated interactions
between membrane proteins are expected to modify the energy cost of bilayer-protein interactions \cite{harroun99,goforth03,botelho06,grage11,nomura12,phillips09,ursell07,pre16}, and hence may affect the thermodynamic competition between membrane protein oligomeric states. In particular, the protein-induced lipid bilayer thickness deformations in eq.~(\ref{eq:elastic_energy}) can yield favorable bilayer-mediated protein interactions $>10$~$k_B T$ in magnitude \cite{phillips09,ursell07,pre16} resulting, as exemplified by MscL \cite{grage11,nomura12}, in clustering of membrane proteins \cite{harroun99,goforth03,botelho06,phillips09}. The lattice architecture of MscL clusters is predicted \cite{scirep16} to depend on the oligomeric state of MscL, with face-on square lattices and distorted hexagonal lattices \cite{henley86} (see fig.~\ref{fig5}(a), central and right insets) providing the ground states of clusters of tetrameric \cite{liu09} and pentameric \cite{chang98} MscL, respectively.

To assess to what extent bilayer-thickness-mediated protein clustering 
\cite{harroun99,goforth03,botelho06,grage11,nomura12,phillips09,ursell07,pre16,scirep16}
may modify the thermodynamic competition between membrane protein oligomeric states, we examine the stability of membrane protein lattices to thermal fluctuations. For a lattice of $n$ membrane proteins in oligomeric state $s$, the bilayer thickness deformation energy in eq.~(\ref{eq:elastic_energy}) per protein in the lattice, $G_{s,n}^{\text{latt}}$, is expected \cite{scirep16} to scale~as
\begin{align} \label{eq:lattice}
 G_{s,n}^{\text{latt}} \sim G^{\infty}_{s} + \frac{G^\prime_s}{\sqrt{n}} \,,
\end{align}
where $G^{\infty}_{s}$ is the energy per protein in infinite lattices and $G^{\prime}_s$ captures boundary effects. Computing the lattice energies of tetrameric \cite{liu09} and pentameric \cite{chang98} MscL clusters in their respective ground-state lattice architectures \cite{scirep16} with the full multi-body interactions implied by eq.~(\ref{eq:elastic_energy}), we find that eq.~(\ref{eq:lattice}) captures the size dependence of $G_{s,n}^{\text{latt}}$ for tetrameric as well as pentameric MscL lattices (see fig.~\ref{fig5}(a)).

Equation~(\ref{eq:lattice}) allows us \cite{benshaul1994,safran2003,bruinsma03} to determine the number of monomers in a lattice of $n$ membrane proteins in oligomeric state $s$, which we denote by $N_{s,n}$, in thermal equilibrium. While we allow here for general cluster sizes, we assume that membrane proteins of distinct oligomeric state form separate clusters with a fixed (ground-state) lattice architecture. We construct the Helmholtz free energy associated with different sizes of (non-interacting) membrane protein lattices following similar steps as for eq.~(\ref{eq:free}), resulting in
\begin{equation}
 F^\text{latt}= \sum_{s,n} \left[N_{s,n} \epsilon_{s,n} +k_B T \frac{N_{s,n}}{s n}\left(\ln \frac{N_{s,n}}{s n N_l}-1 \right) \right] ,
 \label{eq:Fcluster}
\end{equation}
where the energy per monomer is now given by $\epsilon_{s,n}=G^\text{latt}_{s,n}/s$.
Minimization of $F^\text{latt}$ in eq.~(\ref{eq:Fcluster}) with respect to $N_{s,n}$ yields the equilibrium conditions
\begin{align} \label{eq:EqCluster}
\epsilon_{s,n}+ \frac{k_B T}{sn}\ln\frac{N_{s,n}}{sn N_l} =\mu \,,
\end{align}
where $\mu$ is determined by the constraint $\sum_{s,n} N_{s,n}/N_l = M$ fixing the total monomer number fraction in the membrane. Similarly as for eq.~(\ref{eq:distribut}), we numerically solve the set of equations in eq.~(\ref{eq:EqCluster}) to determine the monomer fractions $\phi_{s,n}=N_{s,n}/\sum_{s,n} N_{s,n}$ in clusters of size $n$ composed of membrane proteins of oligomeric state $s$.

Figure~\ref{fig5}(b) shows the monomer fractions $\phi_{s,n}$ obtained from the ground-state lattice architectures \cite{scirep16} of tetrameric \cite{liu09} and pentameric \cite{chang98} MscL clusters (see fig.~\ref{fig5}(a), central and right insets) with a total monomer number fraction $M\approx4.7\times10^{-4}$ corresponding to MscL in \textit{E. coli} under physiological conditions \cite{bialecka12}. We find that almost all MscL subunits in the membrane form tetrameric MscL \cite{liu09}, with $>99\%$ of tetrameric MscL localized in lattices with $N_P-14\leq n \leq N_P$ for a MscL copy number in the cell membrane $N_P\approx580$ associated with $M\approx4.7\times10^{-4}$
\cite{bialecka12}. The dominance of the largest possible cluster sizes in fig.~\ref{fig5}(b) suggests that, to a good approximation, thermal effects can be neglected when examining the competition between tetrameric and pentameric MscL lattices due to bilayer-protein interactions. The dominance of tetrameric over pentameric MscL lattices then follows because tetrameric MscL lattices yield \cite{scirep16} a lower energy per monomer than pentameric MscL lattices (see fig.~\ref{fig5}(a), left inset). However, these conclusions are based on the assumption that membrane proteins can interconvert between different oligomeric states not only in the case of individual, non-interacting membrane proteins \cite{reading15}, but also for close-packed membrane protein lattices \cite{scirep16,grage11}. This might not be a realistic assumption, suggesting that the MscL oligomeric state is fixed before clustering sets in. Indeed, experiments indicate \cite{sukharev99,gandhi11,reading15,walton15,dorwart10,iscla11,haswell11}
that MscL predominantly occur in the pentameric, rather than the tetrameric,
state. Thus, at least in the case of MscL, clustering may not play a dominant role in setting the membrane protein oligomeric~state.

\section{Conclusion}
Based on previous work on amphiphile \cite{benshaul1994,safran2003} and viral capsid \cite{bruinsma03} self-assembly, we have formulated a general physical
framework describing the thermodynamic competition between different oligomeric states of membrane proteins. For MscL we find that, if MscL subunits can either form the observed tetrameric \cite{liu09} or pentameric \cite{chang98} MscL structure with no interactions between MscL, bilayer-protein interactions yield, consistent with experiments \cite{sukharev99,gandhi11,reading15,walton15,dorwart10,iscla11,haswell11}, pentameric MscL as the dominant MscL oligomeric state for physiologically
relevant MscL numbers in the cell membrane \cite{bialecka12}. In contrast, if MscL subunits can also form hexameric MscL shapes \cite{saint98}, hexameric MscL dominate, while direct protein-protein interactions between MscL subunits
and bilayer-mediated interactions between MscL may yield lower-order oligomeric states. Our thermodynamic model suggests that, as also indicated by recent experiments \cite{reading15,walton15}, MscL generally occur as a mixture of different oligomeric states. Theoretical studies have shown \cite{plos13,epl14,scirep16} that distinct oligomeric states of MscL result in distinct gating tensions. Mixtures of MscL oligomeric states in the cell membrane may therefore facilitate a staggered response to changes in membrane tension, helping to increase the functional diversity of mechanosensitive ion channels \cite{perozo03,booth07,kung10,haswell11}.
Our results illustrate how self-assembly of protein monomers into distinct membrane protein oligomeric states \cite{vinothkumar2010,goodsell2000,forrest15,takamori06,yun11,linden12}
may provide a general mechanism for regulation of protein function.

\acknowledgments
This work was supported by NSF award numbers DMR-1554716 and DMR-1206332, an Alfred P. Sloan Research Fellowship in Physics, the James H. Zumberge Faculty Research and Innovation Fund at the University of Southern California, and the USC Center for High-Performance Computing. We also acknowledge support through the Kavli Institute for Theoretical Physics, Santa Barbara, via NSF award number PHY-1125915.


\end{document}